\documentclass[11pt]{article}
\usepackage{amsmath,amssymb}
\usepackage{float}
\usepackage{caption2}
\usepackage{graphicx}
\def\be{\begin{equation}}
\def\ee{\end{equation}}
\def\ba{\begin{eqnarray}}
\def\ea{\end{eqnarray}}

\begin{document}
\title{\textbf{\large{Spectrum of Curvature Perturbation of Multi-field Inflation with Small-Field Potential}}}
\vspace{1cm}
\author{Iftikhar Ahmad\thanks{e-mail:iftikharwah@gmail.com},~~ Yun-Song
Piao\thanks{e-mail:yspiao@gucas.an.cn}~~and~~ Cong-Feng Qiao\thanks{e-mail:qiaocf@gucas.an.cn}\\
\textit{College of physical sciences,Graduate University of Chinese
Academy of Sciences,} \\ \textit{YuQuan Road 19A,Beijing
100049,China.}}

\date{}
\maketitle
\begin{abstract}
In this paper, we have studied the spectrum of curvature
perturbation of multi-field inflation with general small-field
potential. We assume that the isocurvature perturbation may be
neglected, and by using the Sasaki-Stewart formalism, we found
that the spectrum may be redder or bluer than of its corresponding
single field. The result depends upon the values of fields and
their effective masses at the horizon-crossing time. We discuss
the relevant cases.

\vspace{.5cm}\noindent
\end{abstract}
\section{Introduction}
\noindent

Inflation \cite{Guth81, LAS, Star80} naturally solves the
flatness, homogeneity and monopole problems and predicts almost
scale invariant density perturbations, consistent with present
observations. Thus inflation has become the dominant paradigm for
understanding the features of our observable universe. However,
single field inflation model generally has some fine tuning
problems of the parameters, such as the mass and coupling of
field, and also the value of field, which renders it difficultly
be realized in a realistic high energy theory.

When many fields are included, it was found that they can work
cooperatively to drive a period of inflation by assisted inflation
mechanism proposed by Liddle et al. \cite{LMS}, even if any one of
those fields is not able to sustain inflation  separately. The
multi-field inflation model relaxes the difficulties suffered by
single field inflation models, and thus may be regarded as an
attractive implementation of inflation. The assisted inflation
with massive fields have been discussed earlier in \cite{MW}, and
subsequently explored in the Kaluza-Klein model \cite{KO} and in
the Randall-Sundrum model \cite{Piao01} where there would be a
tower of mass eigenstates. Recently, there have been many
interesting examples of multi-field inflation e.g. see
\cite{Piao0206}, in which exponentially large number of field was
required for a feasible theoretic realization of inflation. In
addition, some multi-field inflation models have been also
discussed in \cite{MORE}.

It was shown \cite{DKMW} that the many axion fields predicted by
string vacuum can be combined and lead to a radiatively stable
inflation, called N-flation. These axion fields generally have
different masses, which can be very densely spaced. The spectrum of
curvature perturbation of multi-field inflation with unequal masses
have been discussed in \cite{AL}, and then the detailed study was
made by Easther and McAllister \cite{EM} for quite specific choices
of initial conditions for the fields, and the numerical
investigation was made by Kim and Liddle \cite{KL} for the random
initial conditions. Recently, it was shown analytically in
\cite{YSP} that for multi-field inflation with power law potential,
the spectrum of curvature perturbation is generally redder than that
of its corresponding single field. The result obtained from that
with unequal massive fields is consistent with the numerical
investigation of Kim and Liddle, and is also compatible with recent
other studies \cite{others, KL1}.

It is interesting to generalize above studies to multi-field
inflation with other potentials. The small-field inflation model
arises naturally from spontaneous symmetry breaking such as the
original models of new inflation \cite{ADL,AS}, modular inflation
from string theory \cite{DL} and from Pseudo Nambu-Goldstone modes
(natural inflation) \cite{FFO}. In this class of model, the field
generally starts at near an unstable equilibrium (usually taken to
be at the origin) of its potential and then rolls down along the
potential to a stable minimum. Thus the potential of the small-field
models can be taken as $V(\phi)=\Lambda[1-(\phi/\mu)^{p}]$, which
can be viewed as a lowest-order Taylor expansion of an arbitrary
potential around the origin. A natural generalization to multiple
uncoupled fields may be written as \begin{equation} V= \sum_i
V_i(\phi_i)=\sum_i
\Lambda_i\left[1-(\frac{\phi_i}{\mu_i})^{p}\right],\label{a}
\end{equation}
where the subscript $``i"$ denotes the relevant quantities of the
$i$th field and $p$ is the same for all fields, and $\Lambda_i$ and
$\mu_i$ are the parameters describing the height and tilt of
potential of the $i$th field. In this paper, we will study the
spectrum of curvature perturbation of multi-field inflation with the
potential (\ref{a}).

This article is organized as follows; in section $2$ we calculate
the spectral index of curvature perturbation by using the
Sasaki-Stewart formalism \cite{SS} for $p>2$ and discuss the
relevant results. In section $3$ we focuss on the case of $p=2$.
Finally, in section $4$ we summarize our conclusion.

\section{The Scalar Spectrum of Spectral Index for $ p>2$}
\noindent

With the potential (\ref{a}), in the slow-roll approximation, we
have \be 3H{\dot \phi_i}+V_i^{\prime}(\phi_i) \simeq 0,
\label{phii}\ee \be H^2\simeq {1\over 3M_p^2}\sum_i V_i(\phi_i),
\label{vphi}\ee where $H=\dot a/a$ is Hubble parameter, $a(t)$ is
scalar factor, and $M_{p} =(8{\pi}G)^{-1/2}$ is the Planck mass.
From Eq. (\ref{phii}), we can obtain \be {{\dot \phi}_i\over
V_i^{\prime}(\phi_i)}={{\dot \phi}_j\over V_j^{\prime}(\phi_j)}.
\label{ratio}\ee Thus the efolding number is \ba N & =& \int H dt
=\int H{d\phi_j\over {\dot \phi_j}}\nonumber\\ &\simeq & -\int
3H^2{d\phi_j\over V_j^{\prime}(\phi_j)} \nonumber\\
&\simeq & -{1\over M_p^2}\int \left(\sum_i V_i(\phi_i)\right)
{d\phi_j\over V_j^{\prime}(\phi_j)}\nonumber\\
&=& -{1\over M_p^2}\sum_i\int_{\phi_i^{*}}^{\phi_i^{e}}{V_i\over
V_i^{\prime}} {d\phi_i},\label{b} \ea where Eqs. (\ref{phii}),
(\ref{vphi}) and (\ref{ratio}) have been used, and the upper limit
$\phi_i^e$ of the integral corresponds to the end of inflation and
$``*"$ denotes the horizon-crossing time of corresponding
perturbation and generally $\phi_i^*\lesssim \phi_i^e$. \\
From Eq. (\ref{a}), we have \begin{equation}
\frac{V_i}{V_i^{\prime}}\simeq\frac{1}{p}\left[{\phi_i}-\frac{\mu_i^p}{\phi_i^{p-1}}\right].\label{c}
\end{equation}
Substituting (\ref{c}) into (\ref{b}) and after simplification, we
get
\begin{equation}
N\simeq\frac{1}{pM_p^2}\sum_i\left[\frac{1}{2}({\phi_i^*}^2-{\phi_i^e}^2)+\frac{\mu_i^p}{p-2}
[\frac{1}{{\phi_i^*}^{p-2}}-\frac{1}{{\phi_i^e}^{p-2}}]\right].\label{d}
\end{equation}
Inflation ends at ${\phi_i^e}\lesssim {\mu_i}$, and in the meantime
in the small-field model we have ${\mu_i}\lesssim{M_p}$, thus the
quadratic terms which appear in Eq. (\ref{d}) must vanish since
${\phi_i^*}\lesssim{\phi_i^e}\lesssim{M_p}$, i.e.
~~$(\frac{\phi_i^*}{M_p})^2\rightarrow
0$,~~$(\frac{\phi_i^e}{M_p})^2\rightarrow 0$. Thus we have
\begin{equation}
N\simeq\frac{1}{p(p-2)M_p^2}\sum_i(\frac{\mu_i}{\phi_i^*})^{p-2}\left
[1-(\frac{\phi_i^*}{\phi_i^e})^{p-2}\right]{\mu_i^2}.\label{e}
\end{equation}
Further by taking $1-(\frac{\phi_i^*}{\phi_i^e})^{p-2}\backsimeq 1$,
Eq. (\ref{e}) is reduced to
\begin{equation}
N\simeq\frac{1}{p(p-2)M_p^2}\sum_i(\frac{\mu_i}{\phi_i^*})^{p-2}{\mu_i^2}.
\label{f}~~
\end{equation}

Generally for multi-field there exists not only the curvature
perturbation, but also the orthogonal isocurvature perturbations
\cite{L, GWBM}, see \cite{BTW} for a review. To simplicity, we
assume here that the isocurvature perturbation may be neglected,
which will be further discussed in section $4$. In this case, the
curvature perturbation of multi-field inflation can be evaluated
analytically by using the Sasaki -Stewart formulism \cite{SS} also
earlier in Ref. \cite{AAS}. Thus the spectral index is given by
\begin{equation}
n_s-1\simeq-M_p^2\frac{\sum_i(V_i^{\prime})^2}{V^2}-2M_p^2\frac{1}{\sum_i
(\frac{V_i}{V_i^{\prime}})^2}+2M_p^2\frac{1}{V}\frac{\sum_i(\frac{V_i}{V_i^{\prime}})^2V_i^{\prime\prime}}
{\sum_j(\frac {V_j}{V_j^{\prime}})^2}.\label{g}
\end{equation} To make the relevant formula clear, we introduce a
suitable substitution for summation terms. We introduce
\begin{equation}
A_1=\sum_i(w_i)^{p-2}{\mu_i^2},\label{h}
\end{equation}
where $w_i=\mu_i/\phi_i^*$. Substituting Eq. (\ref{h}) into Eq.
(\ref{f}), the efolding number can be rewritten as
\begin{equation}
N\simeq\frac{A_1}{p(p-2)M_p^2},\label{i}
\end{equation} which implies that
\begin{equation}
M_p^{2}\simeq\frac{A_1}{p(p-2)}\frac{1}{N},\label{j}
\end{equation}
Then we calculate all terms in right side of Eq. (\ref{g})
separately,
\begin{equation}
V^2 \simeq [\sum_i\Lambda_i]^2\equiv A_2^2,\label{k}
\end{equation}
\begin{equation}
\sum_i(V_i^{\prime})^2={p^2}\sum_i
\frac{\Lambda_i^2}{(w_i)^{2(p-1)}\mu_i^2}\equiv {p^2}{A_3},\label{l}
\end{equation}
\begin{equation}
\sum_i(\frac{V_i}{V_i^{\prime}})^2\simeq{1\over
p^2}\sum_i(w_i)^{2(p-1)}\mu_i^2 \equiv\frac{A_4}{p^2},\label{m}
\end{equation}
\begin{equation}
\sum_i(\frac{V_i}{V_i^{\prime}})^2 V_i^{\prime\prime} \simeq
-{(p-1)\over p}\sum_i \Lambda_i(w_i)^p \equiv -{(p-1)\over
p}A_5.\label{n}
\end{equation}
Substituting the above Eqs. into Eq. (\ref{g}), we have the scalar
spectral index
\begin{equation}
n_s-1\simeq-\frac{2}{N}\left(\frac{p-1}{p-2}\right)
\left[\frac{p}{2(p-1)}\left(\frac{A_1A_3}{A_2^2}\right)+\frac{p}{p-1}\left(\frac{A_1}{A_4}\right)
+\frac{A_1A_5}{A_2A_4}\right],\label{p}
\end{equation}
where Eq. (\ref{j}) has been used. Note that
$(1/w_i^{2(p-1)})\simeq 0,\,\,{\rm and}\,\,1/w_i^p\simeq 0$, thus
the first and second terms on right hand side of Eq. (\ref{p})
must vanish, i.e.
$\frac{p}{2(p-1)}\left(\frac{A_1A_3}{A_2^2}\right)\simeq 0,\,\,
{\rm and}\,\,\frac{p}{p-1}\left(\frac{A_1}{A_4}\right)\simeq 0.$
Then we get the spectral index
\begin{equation}
n_s-1\simeq-{2\over N}{{p-1}\over {p-2}}\left[1+
R(w_i)\right],\label{s}
\end{equation}\ where
\begin{equation}
R(w_i)=\frac{(A_1A_5-A_2A_4)}{A_2A_4}.\label{t}
\end{equation} Thus with the definitions of $A_1$, $A_2$, $A_4$ and
$A_5$, we have
\begin{equation}
R(w_i)=\frac{\sum_{i<j}(w_j^p-w_i^p)(\Lambda_jw_i^{p-2}\mu_i^2-\Lambda_iw_j^{p-2}\mu_j^2)}
{\sum_{i,j}\Lambda_iw_j^{2p-2}\mu_j^2}.\label{v}
\end{equation}
The mass $m_i^2$ of each scalar field $\phi_i$ can be defined as
\begin{equation}
m_i^2=V_i^{\prime\prime}=-p(p-1)\Lambda_i\frac{1}{w_i^{p-2}\mu_i^2}.\label{w}
\end{equation}
Thus by substituting Eq. (\ref{w}) into Eq. (\ref{v}), we can
rewrite $R(w_i)$ as
\begin{equation}
R(w_i)=\frac{\sum_{i<j}\Lambda_i\Lambda_j(w_j^p-w_i^p)(\frac{1}{m_i^2}-\frac{1}{m_j^2})}
 {\sum_{i,j}\Lambda_i\Lambda_jw_j^p/m_j^2}.\label{y}
\end{equation}
The first term in the right hand side of Eq. (\ref{s}) is just the
result of single field inflation  \cite{AL}, while the second term
is additional term which appears due to multi-field as shown in
Eq. (\ref{y}). When all fields have same masses at the
horizon-crossing time, we have $R(w_i)=0$, and thus
\begin{equation}
n_s-1\simeq-\frac{2}{N}\left(\frac{p-1}{p-2}\right). \label{z}
\end{equation} In this case the scalar
spectrum of multi-field will be the same as that of its
corresponding single field \cite{AL}. When $w_1<w_2<w_3...<w_n$ and
$m_1^2<m_2^2<...<m_n^2$, $R(w_i)$ will be always positive, which
means that spectrum is more redder than that of its corresponding
single field. But when $w_1<w_2<w_3...<w_n$ and
$m_1^2>m_2^2>...>m_n^2$, $R(w_i)$ will be always negative, which
means that the spectrum is less red than that of its corresponding
single field. However, for more general cases, it seems that
dependent of parameters of fields and initial conditions, there is
no definite conclusion.

In the case of two scalar fields,  from Eq. (\ref{y}), we can get
\begin{equation}
R(w_1,w_2)=\frac{\Lambda_1\Lambda_2(w_2^p-w_1^p)(1/m_1^2-1/m_2^2)}
{(\Lambda_1+\Lambda_2)(\Lambda_1 w_1^p/m_1^2+\Lambda_2
w_2^p/m_2^2)}\simeq
\frac{(w_2^p-w_1^p)(m_2^2-m_1^2)}{(1/\Lambda_2+1/\Lambda_1)(\Lambda_1
m_2^2w_1^p+\Lambda_2 m_1^2w_2^p)},\label{aa}
\end{equation}
if $w_1<w_2$ and $m_1^2<m_2^2$, $R(w_1,w_2)$ will be positive,
which implies that the  scalar spectrum is more red than that of
its corresponding single field, but if $w_1<w_2$ and
$m_1^2>m_2^2$, $R(w_1,w_2)$ will be negative, which implies that
the scalar spectrum is less red than that of its corresponding
single field.

 \section{The Scalar Spectrum  for $p=2$}
 \noindent

The Eq. (\ref{s}) is only valid for $p>2$, so we need to separately
calculate the case of $p=2$ which will be done in this section. We
consider a general potential as
 $ V=\sum_i \Lambda_i \left[1-(\frac{\phi_i}{\mu_i})^2\right]$. In
 this case
$V_i^{\prime}={-2} \Lambda_i\frac{\phi_i}{\mu_i^2}$ and
  $V_i^{\prime\prime}={-2} \Lambda_i\frac{1}{\mu_i^2}$. Moreover,
  we can approximate $(\frac{\phi_i^*}{M_p})^2\longrightarrow 0$,
 $(\frac{\phi_i^e}{M_p})^2\longrightarrow
 0$. Thus, following the same steps
as we did in the previous section, we obtain
 \begin{equation}
 N\simeq-
 \frac{1}{2M_p^{2}}\sum_i\ln\left(\frac{\phi_i^*}{\phi_i^e}\right)\mu_i^2.\label{bb}
\end{equation}
From Eq. (\ref{g}), we get the spectral index
\begin{equation}
n_s-1\simeq-2M_p^2\left[2\frac{\sum_i(\Lambda_i^2/{\mu_i^2
w_i^2})}{(\sum_i \Lambda_i)^2}+4\frac{1}{\sum_i\mu_i^2
w_i^2}+2\frac{\sum_i \Lambda_i w_i^2}{\sum_i\Lambda_i\sum_j\mu_j^2
w_j^2}\right].\label{gg}
\end{equation}
Note that $1/w_i^2\simeq 0$, thus the first two terms on right side
of above equation can be neglected, which reduces Eq. (\ref{gg}) to
\begin{equation}
 n_s-1\simeq-4 M_p^2\frac{\sum_i \Lambda_i
w_i^2}{\sum_i\Lambda_i\sum_j\mu_j^2 w_j^2}.\label{hh}
\end{equation}
From Eq. (\ref{ratio}), we have
\begin{equation}
\frac{\dot\phi_i}{\dot\phi_j}\simeq \frac{\Lambda_i
\phi_i/\mu_i^2}{\Lambda_j \phi_j/\mu_j^2},\label{kk}
\end{equation} which implies that
\begin{equation}
\mu_i^2\frac{\dot\phi_i}{\Lambda_i \phi_i}\simeq
\mu_j^2\frac{\dot\phi_j}{\Lambda_j \phi_j}.\label{ll}
\end{equation}
By taking the integral on both side of Eq. (\ref{ll}), we get
\begin{equation}
\frac{\mu_i^2}{\Lambda_i}\ln(\frac{\phi_i^*}{\phi_i^e})=
\frac{\mu_j^2}{\Lambda_j}\ln(\frac{\phi_j^*}{\phi_j^e}),\label{mumu}
\end{equation} which implies that
\begin{equation}
\sum_i\mu_i^2\ln(\frac{\phi_i^*}{\phi_i^e})=\mu_k^2
\ln(\frac{\phi_k^*}{\phi_k^e})\sum_i\frac{\Lambda_i}{\Lambda_k}~~~~~
{\rm by~ taking~} j=k, \label{nn}
\end{equation} It should be noted that for the fixed value of ``k" we get relation (\ref{nn}),
in which the term with subscript ``k" is independent of summation
notation. Now substituting Eq. (\ref{nn}) into Eq. (\ref{bb}), we
obtain
\begin{equation}
M_p^2\simeq-\frac{1}{2N}\mu_k^2\ln(\frac{\phi_k^*}{\phi_k^e})\sum_i\frac{\Lambda_i}{\Lambda_k}\label{mm}.
\end{equation} which then is substituted into Eq. (\ref{hh}) and gives
\begin{equation}
 n_s-1\simeq\frac{2}{N}\ln(\frac{\phi_k^*}{\phi_k^e})\frac{\sum_i
 w_i^2\left(\frac{\Lambda_i}{\Lambda_k}\right)}{\sum_j (\mu_j^2/\mu_k^2) w_j^2}.
\label{ooo}
\end{equation}
If all fields have same $\Lambda_i$, we will obtain
\begin{equation}
n_s-1\simeq\frac{2}{N}\ln(\frac{\phi_k^*}{\phi_k^e})\frac{\sum_i
w_i^2}{\sum_j (\mu_j^2/\mu_k^2) w_j^2}.\label{oo}
\end{equation}
The results of Eqs. (\ref{ooo}) and (\ref{oo}) are strictly
independent of the choice of $k$. This can be seen as follows. For
example, we take $k=k_1$ and $k=k_N$, and put them into Eq.
(\ref{oo}). If we call the corresponding indices $n_{k_1}-1$ and
$n_{k_N}-1$ respectively, then ${(n_{k_1}-1)/(n_{k_N}-1)}$ is
simply $1$ because of Eq. (\ref{mumu}) with $\Lambda_i$ being the
same for all $i$.

Further, if all fields have same $\mu_i$, we will have
\begin{equation}
n_s-1\simeq\frac{2}{N}\ln(\frac{\phi_k^*}{\phi_k^e}). \label{pp}
\end{equation}
Note that in this case all $\ln(\frac{\phi_i^*}{\phi_i^e})$ are
equivalent due to Eq. (\ref{mumu}). Eq.(\ref{pp}) is actually same
as with that for its corresponding single field.

Now let us see Eq. (\ref{oo}). If all $\Lambda_i$ are taken to be
same while $\mu_i$ are different, from Eq. (\ref{mumu}), it is
obvious that a larger $\mu_i$ will correspond to a smaller
$\ln(\frac{\phi_i^*}{\phi_i^e})$. When taking $\mu_k={\rm
Max}(\mu_1,\mu_2,...\mu_n)$, which implies that $\mu_i/\mu_k<1$,
from Eq. (\ref{oo}), we have the spectrum is more red than that of
single field $\phi_k$, i.e. Eq. (\ref{pp}). Note that in this case
$\ln(\frac{\phi_k^*}{\phi_k^e})$ is the smallest in all
$\ln(\frac{\phi_i^*}{\phi_i^e})$, which suggests that in term of Eq.
(\ref{pp}), the spectrum for single field $k$ is the nearest from 1.
When taking $\mu_k={\rm Min}(\mu_1,\mu_2,...\mu_n)$, which implies
that $\mu_i/\mu_k>1$, from Eq. (\ref{oo}), we have the spectrum is
less red than that of single field $\phi_k$. Note that in this case
$\ln(\frac{\phi_k^*}{\phi_k^e})$ is the largest in all
$\ln(\frac{\phi_i^*}{\phi_i^e})$. The above results means that for
$p=2$ the scalar spectral index generally lies between that of
single field with the largest $\mu_k$ and that of single field with
the smallest $\mu_k$.

\section{Conclusion}

We have studied the spectrum of curvature perturbation of
multi-field inflation with general small-field potential. For
$p>2$, we found that the spectrum may be redder or bluer than that
of its corresponding single field. The result is depending on the
value of fields and their effective masses at the horizon-crossing
time. This result is different from that of multi-field inflation
with power law potential, in which the definite conclusion that
the spectrum is redder than that of its corresponding single field
may be obtained. When the effective masses of all fields are
equal, the spectrum will be the same with its corresponding single
field. The behavior for $p=2$ is different from that of $p>2$. In
this case we observed that the scalar spectral index generally
lies between that of single field with the largest $\mu_k$ and
that of single field with the smallest $\mu_k$. When all $\mu_i$
are taken to be equivalent, the spectrum will be the same with
that of its single field.

In our study, the results are dependent of not the initial
conditions, but the values of fields and their effective masses at
the horizon-crossing time, since the perturbation spectrum is
calculated at the horizon-crossing time of corresponding
perturbation, which in fact is a good approximation, as was
discussed in Ref. \cite{KL}. In order to obtain this result, we
only require that the slow roll approximation for each field is
satisfied initially and also at all time.

It should be noted that here we have assumed that the isocurvature
perturbation may be neglected. In fact, in order to neglect the
isocurvature perturbation, the conditions ${\delta\phi_i/ {\dot
\phi}_i}={\delta\phi_j/ {\dot \phi}_j}$ have to be imposed. In this
case, it seems that the fields must be constrained in some special
trajectories. In addition, it is also possible that for general
trajectories at some special horizon-crossing time of corresponding
perturbation, the above condition is just satisfied. The results
here are actually suitable for above both cases. Though in general
case the inclusion of isocurvature mode seems required, such study
for general large N fields is beyond our ability at present.
However, our work may be regarded as one interesting step along this
line.

\section{Acknowledgment}
I.A would like to thanks the  support of (HEC)Higher Education
Commission of Pakistan in favor of research in China. This work is
supported in part by the National Natural Science Foundation of
China (NSFC) No: 10491306, 10521003, 10775179, 10405029, 10775180,
in part by the Scientific Research Fund of GUCAS(NO.055101BM03),
in part by CAS under Grant No: KJCX3-SYW-N2.

\end{document}